\documentclass[prc,twocolumn,showpacs,preprintnumbers,amsmath,amssymb]{revtex4}
\usepackage{txfonts}
\usepackage{amssymb}
\usepackage{graphicx}

\usepackage{tipa}
\usepackage{dcolumn}
\usepackage{bm}
\usepackage{multirow}
\usepackage{booktabs}

\begin{document}
\title{Beta-decay study of $T_z=-2$ proton-rich nucleus $^{20}$Mg}
\author{L. J. Sun$^{1}$}
\author{X. X. Xu$^{1}$}
 \email{xuxinxing@ciae.ac.cn}
\author{C. J. Lin$^{1}$}
 \email{cjlin@ciae.ac.cn}
\author{J. S. Wang$^{2}$}
\author{D. Q. Fang$^{3}$}
\author{Z. H. Li$^{4}$}
\author{Y. T. Wang$^{3}$}
\author{J. Li$^{4}$}
\author{L. Yang$^{1}$}
\author{N. R. Ma$^{1}$}
\author{K. Wang$^{3}$}
\author{H. L. Zang$^{4}$}
\author{H. W. Wang$^{3}$}
\author{C. Li$^{3}$}
\author{C. Z. Shi$^{3}$}
\author{M. W. Nie$^{3}$}
\author{X. F. Li$^{3}$}
\author{H. Li$^{3}$}
\author{\\J. B. Ma$^{2}$}
\author{P. Ma$^{2}$}
\author{S. L. Jin$^{2}$}
\author{M. R. Huang$^{2}$}
\author{Z. Bai$^{2}$}
\author{J. G. Wang$^{2}$}
\author{F. Yang$^{1}$}
\author{H. M. Jia$^{1}$}
\author{H. Q. Zhang$^{1}$}
\author{Z. H. Liu$^{1}$}
\author{P. F. Bao$^{1}$}
\author{D. X. Wang$^{1}$}
\author{Y. G. Ma$^{3}$}
\author{Y. Y. Yang$^{2}$}
\author{Y. J. Zhou$^{2}$}
\author{W. H. Ma$^{2}$}
\author{J. Chen$^{2}$}

\affiliation{$^1$Department of Nuclear Physics, China Institute of Atomic Energy, Beijing 102413, People's Republic of China\\
$^2$Institute of Modern Physics, Chinese Academy of Sciences, Lanzhou 730000, People's Republic of China\\
$^3$Shanghai Institute of Applied Physics, Chinese Academy of Sciences, Shanghai 201800, People's Republic of China\\
$^4$State Key Laboratory of Nuclear Physics and Technology, School of Physics, Peking University, Beijing 100871, People's Republic of China}

\date{\today}

\begin{abstract}
The $\beta$ decay of the drip-line nucleus $^{20}$Mg gives important information on resonances in $^{20}$Na, which are relevant for the astrophysical $rp$-process. A detailed $\beta$ decay spectroscopic study of $^{20}$Mg was performed by a continuous-implantation method. A detection system was specially developed for charged-particle decay studies, giving improved spectroscopic information including the half-life of $^{20}$Mg, the excitation energies, the branching ratios, and the log~$ft$ values for the states in $^{20}$Na populated in the $\beta$ decay of $^{20}$Mg. A new proton branch was observed and the corresponding excited state in $^{20}$Na was proposed. The large isospin asymmetry for the mirror decays of $^{20}$Mg and $^{20}$O was reproduced, as well. However, no conclusive conclusion can be draw about the astrophysically interesting 2645~keV resonance in $^{20}$Na due to the limited statistics.
\end{abstract}

\pacs{23.50.+z, 23.40.-s, 23.20.Lv, 27.30.+t}
\keywords{$\beta$-delayed proton emission; Double-sided silicon strip detector; Proton spectrum; Half-life; Decay branching ratio}
\maketitle

\section{Introduction}
Information about the short-lived nuclei are essential to revealing some of the long-standing mysteries of the astrophysical rapid proton capture process ($rp$-process) \cite{Schatz_PR1998}, namely, the main astronomical site and its mechanism. $\beta$ decay can be a good way to study some specific resonances in the daughter nucleus under some stellar environments, for example, determine the spin and parity of the resonances populated in the $\beta$ decay on the basis of the selection rules. A large number of decay channels including $\beta$-delayed particle emission will open due to the high $\beta$ decay energy and low separation energy of nucleons at the drip-lines. Studies of the $\beta$ decay and $\beta$-delayed particle emission of exotic nuclei also advanced our understanding of the nature of the basic interactions which affect the structure of the nucleus \cite{Blank_PPNP2008,Borge_PS2013,Pfutzner_RMP2012}.

The $^{19}$Ne($p,\gamma$)$^{20}$Na reaction is part of the breakout sequence from the hot CNO cycle, as $^{20}$Na is located on the onset of $rp$-process. The reaction rate of ($p,\gamma$) reaction is dependent on resonance energies and resonance strengths ($\omega\gamma$). For the  $^{19}$Ne($p,\gamma$)$^{20}$Na reaction of astrophysical interest, its reaction rate is dominated by low-energy resonant levels in $^{20}$Na. The nuclear properties (excitation energy, spin and parity, partial decay widths) of the states near and just above the proton-separation threshold in $^{20}$Na play a key role to estimate the reaction rate. In particular, the first excited state above the threshold was found at 2645~keV in $^{20}$Na, whereas its property has been controversial for 30~years \cite{Wiescher_JPG1999,Kubono_NPA1995}. The resonance energy was well-known but only an upper limit of the resonance strength was obtained. A series of experimental and theoretical studies through the $^{20}$Ne($^{3}\mathrm{He},t$)$^{20}$Na charge exchange reaction \cite{Lamm_ZPA1987,Lamm_NPA1990,Kubono_ZPA1988,Kubono_APJ1989,Clarke_JPG1990,Clarke_JPG1993,Smith_NPA1992,Gorres_PRC1995,Hofstee_NIMB1995,Anderson_PRC1995}, the $^{20}$Ne($p,\gamma$)$^{20}$Na reaction with radioactive $^{19}$Ne beams via inverse-kinematics method \cite{Huyse_NPA1995,Huyse_JPG2011,Coszach_PRC1994,Page_PRL1994,Michotte_PLB1996,Vancraeynest_NPA1997,Vancraeynest_PRC1998,Couder_PRC2004}, shell-model calculations \cite{Brown_PRC1993,Fortune_PRC2000}, and various other approaches \cite{Gorres_PRC1994,Seweryniak_PLB2004,Wallace_PRC2012} have been conducted. However, the spin and parity assignment of the state is unsettled even after these extensive investigations. The two most likely spin and parity for this state are 1$^{+}$ and 3$^{+}$. The state can be populated in the allowed transition of $^{20}$Mg in the former case while in the latter case the transition will be strongly forbidden \cite{Wallace_PLB2012}. The $\beta$ decay of $^{20}$Mg can be used as an alternative way to investigate the configuration of the 2645~keV state in $^{20}$Na.
Apart from the measurement of the decay properties of the resonances populated in the $\beta$ decay of $^{20}$Mg, other motivations for studying the $\beta$ decay of the lightest bound magnesium isotope, i.e., $^{20}$Mg are to measure the $\beta$ decay strength distribution and investigate the quenching of Gamow-Teller strength in $\beta$ decay, to test the isobaric multiplet mass equation and to study the isospin symmetry in comparison with the mirror decay and the mirror nucleus \cite{Piechaczek_NPA1995}.

The $\beta$ decay study of $T_{z}=-2$ proton-rich nucleus $^{20}$Mg has been performed with various detection methods. The $\beta$-delayed protons from $^{20}$Mg decay were first observed through helium-jet techniques by D. M. Moltz \textit{et al}. \cite{Moltz_PRL1979} in 1979, providing the first test of the validity of the isobaric multiplet mass equation for the $A=20$ quintet in spite of the low statistics and the high contamination from $^{20}$Na. In 1992,  two $\beta$-delayed proton spectroscopic studies of $^{20}$Mg were performed by S. Kubono \textit{et al}. \cite{Kubono_NIMB1992,Kubono_PRC1992} and J. G\"{o}rres \textit{et al}. \cite{Gorres_PRC1992}, respectively. Both of them implanted the projectile fragments into silicon detectors and more $\beta$-delayed proton peaks from $^{20}$Mg decay were observed. S. Kubono \textit{et al} estimated an upper limit of 1\% for the branching ratio to the 2637~keV state in $^{20}$Na and assigned this state to be the analog of the 3175~keV 1$^+$ state in $^{20}$F, while J. G\"{o}rres \textit{et al} reduced this upper limit to 0.2\%. The most comprehensive $\beta$ decay spectroscopy of $^{20}$Mg was performed by A. Piechaczek \textit{et al}. \cite{Piechaczek_NPA1995} in 1995. Both of the protons and $\gamma$-rays were measured, from which an improved decay scheme of $^{20}$Mg was constructed. An upper limit of 0.1\% for the branching ratio to the 2645~keV state was determined as well. Recently in 2012, J. P. Wallace \textit{et al}. \cite{Wallace_PLB2012} performed a $\beta$-delayed proton spectroscopic study by implanting the ions into a very thin double-sided silicon strip detector. They reported a more stringent upper limit on the branching ratio to the 2647~keV state of 0.02\% with a 90\% confidence level, which strongly supported a 3$^+$ assignment, being the analog of the 2966~keV 3$^+$ state in $^{20}$F. A breakdown of the isobaric multiplet mass equation in the $A=20, T=2$ quintet was reported by A. T. Gallant \textit{et al}. \cite{Gallant_PRL2014}. Soon in 2015, the latest $\beta$ decay study of $^{20}$Mg was done by B. E. Glassman \textit{et al}. \cite{Glassman_PRC2015}. They measured the $\beta$-delayed $\gamma$-rays from $^{20}$Mg decay and determined the excitation energy of the lowest $T=2$ state in $^{20}$Na with high precision. The isobaric multiplet mass equation for the $A=20$ quintet was found to be revalidated.

It is a serious challenge to assign the proton peaks to the right decay branches and reconstruct the decay scheme as numerous states in $^{20}$Na and the proton daughter nucleus $^{19}$Ne are populated in the $\beta$ decay of $^{20}$Mg \cite{Piechaczek_NPA1995}. In the present paper, we report the detailed information about the complicated decay of $^{20}$Mg obtained by measuring the emitted particles and $\gamma$-rays in the $\beta$ decay with high efficiency and high resolution. For the sake of completeness, the preliminary results of an experiment \cite{Lund_PhD2016} performed a few months after the present experiment were also included in this paper.

\section{Experimental techniques}
The experiment was performed at the Heavy Ion Research Facility of Lanzhou (HIRFL) \cite{Zhan_NPA2008} in December 2014. A $^{28}$Si primary beam at 75.8~MeV/nucleon with an intensity of $\sim$37~enA ($\sim$2.6~pnA) impinged on a 1500~$\mathrm{\mu}$m thick $^9$Be target. The main setting of the Radioactive Ion Beam Line in Lanzhou (RIBLL) \cite{Sun_NIMA2003} for the selection of the secondary beam was optimised on $^{22}$Si, and the relevant results will be published elsewhere \cite{Xu_PRL2016}. The ions in the secondary beam were identified by energy-loss ($\mathrm{\Delta}E$) and time-of-flight (ToF) with respect to two focus planes of the RIBLL given by silicon detectors and two scintillation detectors, respectively. In the secondary beam, the accompanying $^{20}$Mg ions were provided with an average intensity of 0.59~particles per second and an average purity of 0.13\%. In order to develop an advanced detection system with high detection efficiency and low detection threshold for charged-particle in the decay, several technologies and solutions were conceived and implemented. Details concerning the detection setup were described in Ref.~\cite{Sun_NIMA2015}, and here we give only the main features. The isotopes of interest were implanted into two double-sided silicon strip detectors (DSSD1 of 149~$\mathrm{\mu}$m thickness and DSSD2 of 66~$\mathrm{\mu}$m thickness), which also served as the subsequent decay detectors. A 314~$\mathrm{\mu}$m thick quadrant silicon detector \cite{Bao_CPC2014} (QSD1) was mounted downstream to serve as an anticoincidence of the penetrating heavy ions and also to detect the charged particles escaping from the DSSD2. A 1546~$\mathrm{\mu}$m thick QSD2 was installed downstream to detect the $\beta$ particles. QSD3 and QSD4, each with a thickness of $\sim$300~$\mathrm{\mu}$m, were installed at the end to suppress the possible disturbances from the penetrating light particles ($^1$H, $^2$H, $^3$H, and $^4$He) coming along with the beam. Besides, the silicon detectors were surrounded by five clover-type HPGe detectors, which were employed to measure the $\gamma$-rays. In front of the silicon detectors array, an aluminum degrader was installed to adjust the stopping range of the ions in the DSSDs. The known $\beta$-delayed protons from $^{21}$Mg decay \cite{Sextro_PRC1973} measured in the previous stage of the experiment were used for the energy calibrations of the DSSDs. The known $\beta$-delayed $\gamma$-rays from $^{22}$Mg decay \cite{Hardy_PRL2003} and $^{24}$Si decay \cite{Ichikawa_PRC2009} measured in the latter stage of the experiment were used for the absolute efficiency calibrations of the clover detectors. The clover detectors were also calibrated in energy and intrinsic efficiency with a $^{152}$Eu standard source.

\section{Results}
\begin{figure}
\begin{center}
\includegraphics[width=2.5in]{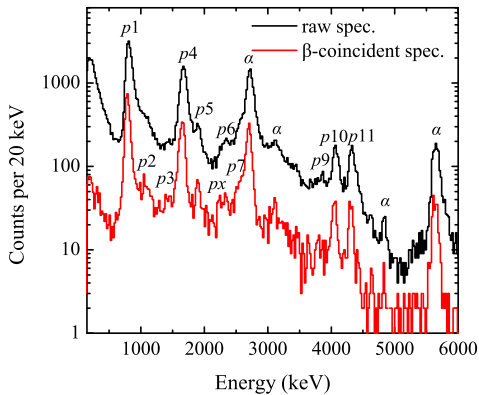}
\caption{\label{ParticleSpec}$\beta$-delayed particle spectra from $^{20}$Mg decay measured by the two DSSDs. The red curve represents the $\beta$-coincident particle spectrum. The proton peaks come from the $\beta$-delayed proton decay of $^{20}$Mg are labeled with ``$p$+numbers" and the $\alpha$ peaks come from the $\beta$-delayed $\alpha$ decay of $^{20}$Na are labeled with letters $\alpha$.}
\end{center}
\end{figure}

The total $\beta$-delayed particle spectrum from $^{20}$Mg decay measured by the two DSSDs is presented in Fig.~\ref{ParticleSpec}. The time difference between an implantation event and all the subsequent decay events was limited within five half-life windows (450~ms). Charged-particles escaping from the DSSD will deposit incomplete energies in the DSSD and the residual energies of the escaping charged-particles can be measured by the other DSSD with high efficiency. Fig.~\ref{ParticleSpec} shows the sum of the deposited energy in DSSD1 and DSSD2. The $\beta$ pile-up effect of the particle spectrum can be reduced by requiring a coincidence with $\beta$ signals in the QSD2. This condition selects the decay events with short flight paths of $\beta$ particles in the DSSDs, hence more proton peaks can be identified in the $\beta$-coincident spectrum. As shown in Fig.~\ref{ParticleSpec}, the $\beta$-delayed protons from $^{20}$Mg decay are marked with ``$p$+numbers" and the $\beta$-delayed $\alpha$ from $^{20}$Na decay are marked with letters $\alpha$. The origin of each particle peak in the spectrum can be identified with the half-life analysis. A new weak peak labeled with $px$ at 2256~keV is observed in the particle spectrum, which is confirmed to be the $\beta$-delayed protons from $^{20}$Mg decay as its half-life is estimated to be $101.9\pm14.9$~ms. In a previous measurement \cite{Wallace_PLB2012}, a $\sim$2340 keV peak was indicated in the $\beta$-delayed particle spectrum from $^{20}$Mg decay. It is to be noted that the shape of their $\sim$2340 keV peak is much broader than those of other peaks, while the relatively better resolution and higher sensitivity achieved in the present work made it possible to clearly distinguish the two peaks unresolved in their $\sim$2340 keV peak. The proton decay branching ratios can be calculated by counting the $\beta$-delayed proton decay events in the particle spectrum, divided by the numbers of the implanted $^{20}$Mg ions. The background subtraction of the proton numbers, the proton detection efficiency correction of the DSSDs and the dead-time correction of the data acquisition system should be applied, as well. The energies and the branching ratios for the $\beta$-delayed protons from $^{20}$Mg decay observed in the present work are summarized in Table~\ref{E20Mg}, and the agreement with the literature values is good within the error for all the proton groups. The errors for energies are attributed to the uncertainties of the calibration parameters and the Gaussian fitting uncertainties of the peak-energies. The errors for branching ratios include the statistical errors and the uncertainties from the background subtraction, the detection efficiency correction and the dead-time correction.

\begin{table*}\scriptsize
\caption{\label{E20Mg}Center-of-mass energies ($E_p$) and branching ratios ($br$) for $\beta$-delayed protons from $^{20}$Mg decay.}
\begin{center}
\begin{ruledtabular}
\begin{tabular}{ccccccccccccc}
    & \multicolumn{2}{c}{S. Kubono \cite{Kubono_PRC1992}} & \multicolumn{2}{c}{J. G\"{o}rres \cite{Gorres_PRC1992}} & \multicolumn{2}{c}{A. Piechaczek \cite{Piechaczek_NPA1995}} & \multicolumn{2}{c}{J. P. Wallace \cite{Wallace_PLB2012}} & \multicolumn{2}{c}{M. V. Lund \cite{Lund_PhD2016}} & \multicolumn{2}{c}{Present Work} \\
    Proton\footnotemark[1] & $E_p$ (keV) & $br$ (\%) & $E_p$ (keV) & $br$ (\%) & $E_p$ (keV) & $br$ (\%) & $E_p$ (keV) & $br$ (\%) & $E_p$ (keV) & $br$ (\%) & $E_p$ (keV) & $br$ (\%) \\
\hline
    $p1$    & 847   & 9     & 807(10) & 10.7(5) & 806(2) & 11.5(14) & 797(2) &       & 780(8) &       & 808(13) & 8.6(7) \\
          &       &       &       &       &       &       & 885(15) & 0.5(1) &       &       &       &  \\
    $p2$    &       &       &       &       & 1056(30) & 0.7(1) &$\sim1050$ &       &       &       & 1071(18) & 0.7(2) \\
    $p3$    &       &       &       &       & 1441(30) & -     &       &       &       &       & 1416(18) & 0.4(1) \\
    $p4$    & 1669  & 5     & 1670(10) & 5.4(5) & 1679(15) & 4.8(6) & 1670(10) &       & 1656(10) &       & 1673(14) & 5.6(5) \\
    $p5$    & 1891  &       &       &       & 1928(16) & 1.1(2) & 1903(5) &       & 1907(3) &       & 1897(17) & 1.1(1) \\
          &       &       &       &       &       &       &       &       & 2138(6) &       &       &  \\
    $px$    &       &       &       &       &       &       &       &       &       &       & 2256(18) & 0.3(1) \\
    $p6$    & 2351  &       &       &       & 2344(25) & 0.3(1)+0.8(1) & $\sim2340$ &       & 2335(3) &       & 2359(18) & 0.4(1) \\
    $p7$    &       &       &       &       & 2559(45) & -     &       &       & 2567(4) &       & 2576(20) & 0.2(1) \\
    $p8$    & 2865  &       &       &       & 2884(45) & -     &       &       & 2768(6) &       &       &  \\
          &       &       &       &       &       &       &       &       & 3081(12) &       &       &  \\
          &       &       &       &       &       &       &       &       & 3320(6) &       &       &  \\
    $p9$    &       &       &       &       & 3837(35) & 0.2(1)+0.1(1) &       &       & 3817(3) &       & 3853(17) & 0.3(1) \\
    $p10$   & 3990  & 0.8   & 4098(19) & 1.3(6) & 4071(30) & 0.7(1)+0.59(1)+0.32(1) & $\sim4080$ &       & 4051(2) &       & 4076(16) & 0.9(1) \\
    $p11$   & 4239  & 0.7   & 4332(16) & 1.7(6) & 4326(30) & 1.8(3) & 4332(16) &       & 4303(4) &       & 4337(16) & 1.0(1) \\
          &       &       &       &       &       &       &       &       & 4544(25) &       &       &  \\
          &       &       &       &       &       &       &       &       & 4993(16) &       &       &  \\
\end{tabular}
\end{ruledtabular}
\footnotetext[1]{The label numbers of proton peaks correspond to the label numbers in Fig.~\ref{ParticleSpec} as well as those in Ref.~\cite{Piechaczek_NPA1995}.}
\end{center}
\end{table*}

\begin{figure}
\begin{center}
\includegraphics[width=2.5in]{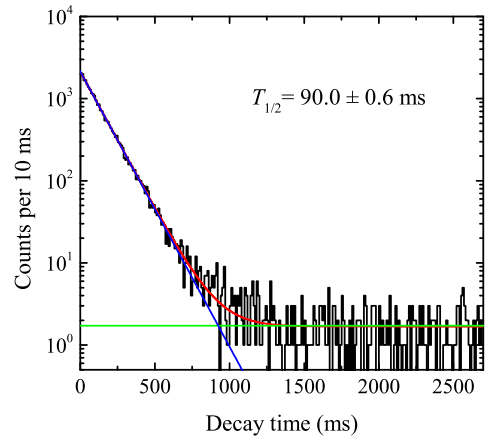}
\caption{\label{TimeSpec}Decay-time spectrum of $^{20}$Mg. The spectrum is fitted with a formula (red line) which can be decoupled into an exponential decay component (blue line) and a constant background component (green line).}
\end{center}
\end{figure}

As shown in Fig.~\ref{TimeSpec}, the decay-time spectrum of $^{20}$Mg is generated by the summation of time difference between an implantation event and all the subsequent decay events which occurs in the same $x$-$y$ pixel of the DSSD. In order to eliminate the influence of the $\beta$-delayed $\alpha$ decay of $^{20}$Na (the daughter of $^{20}$Mg $\beta$ decay), only the two strongest proton peaks ($p1$ and $p4$) are taken into account. The decay-time spectrum contains a small quantity of random correlations, in which the implantation events could be accidentally correlated with decay events from other implantation events or disturbance events from background. All the true correlated implantation and decay event-pairs generate an exponential curve whereas all the uncorrelated event-pairs yield a constant background. In Fig.~\ref{TimeSpec}, a fit with a function composed of an exponential decay and a constant background yields the half-life of $^{20}$Mg to be $90.0\pm0.6$~ms. The uncertainty is derived from the fitting program. The $\chi^2/\mathrm{NDF}=1.14$ represents a good fit, where the ``NDF" refers to the number of degrees of freedom. The result is tabulated and compared with the literature values in Table~\ref{T20Mg}, and nice agreement is obtained.

\begin{table}
\caption{\label{T20Mg}Half-lives of $^{20}$Mg.}
\begin{center}
\begin{ruledtabular}
\begin{tabular}{ccc}
Literature & Year & $T_{1/2}$ (ms)\\
\hline
D. M. Moltz \cite{Moltz_PRL1979} & 1979 & $95^{+80}_{-50}$\\
S. Kubono \cite{Kubono_PRC1992} & 1992 & $114\pm17$\\
J. G\"{o}rres \cite{Gorres_PRC1992} & 1992 & $82\pm4$\\
A. Piechaczek \cite{Piechaczek_NPA1995} & 1995 & $95\pm3$\\
Shell-model calculation \cite{Piechaczek_NPA1995} & 1995 & $101.8$\\
J. P. Wallace \cite{Wallace_PLB2012} & 2012 & $\sim90$\\
M. V. Lund \cite{Lund_PhD2016} & 2016 & $90.9\pm1.2$\\
Present work & 2016 & $90.0\pm0.6$\\
\end{tabular}
\end{ruledtabular}
\end{center}
\end{table}

\begin{figure}
\begin{center}
\includegraphics[width=2.5in]{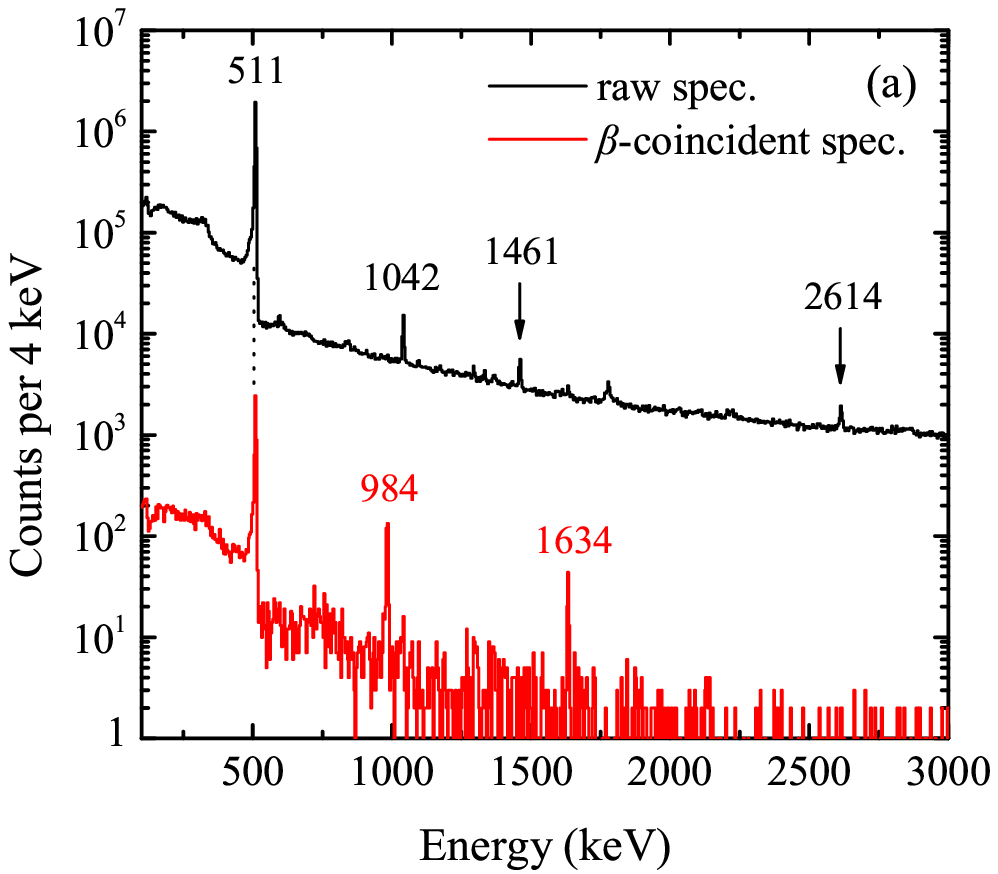}
\includegraphics[width=2.5in]{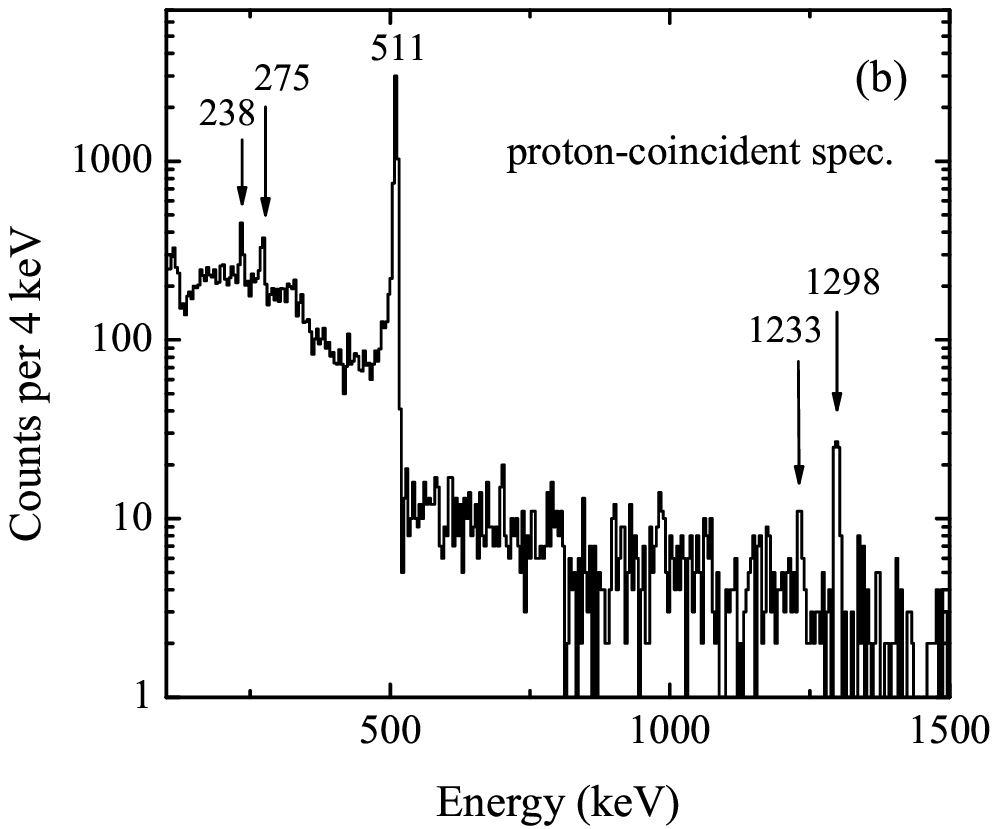}
\caption{\label{GammaSpec}$\gamma$-ray spectrum measured by the clover detectors. (a) the raw $\gamma$-ray spectrum without any coincidence and the $\gamma$-ray spectrum in coincidence with $\beta$ particles from $^{20}$Mg decay. (b) the $\gamma$-ray spectrum in coincidence with charged-particles from $^{20}$Mg decay.}
\end{center}
\end{figure}

\begin{figure}
\begin{center}
\includegraphics[width=2.5in]{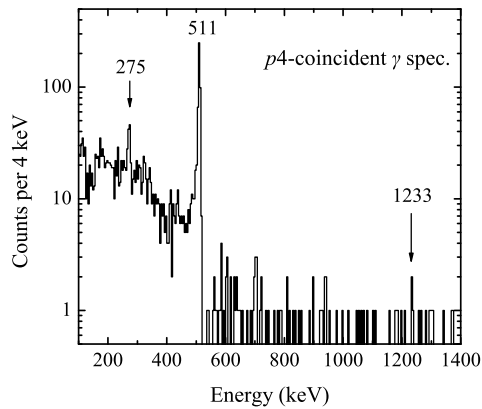}
\caption{\label{p4GammaSpec}$\gamma$-ray spectrum in coincidence with $p4$ from $^{20}$Mg decay measured by the two DSSDs.}
\end{center}
\end{figure}

Fig.~\ref{GammaSpec}(a) shows the raw $\gamma$-ray spectrum without any coincidence and the $\gamma$-ray spectrum in coincidence with $\beta$ particles from $^{20}$Mg decay. Fig.~\ref{GammaSpec}(b) shows the $\gamma$-ray spectrum with coincidence gating condition on charged-particles from $^{20}$Mg decay. Until now, the only $\gamma$-ray measurements of $^{20}$Mg decay were conducted by A. Piechaczek \textit{et al}. at GANIL \cite{Piechaczek_NPA1995} and recently by B. E. Glassman \textit{et al}. at NSCL \cite{Glassman_PRC2015}. In Fig.~\ref{GammaSpec}(a), the 984~keV $\gamma$-ray comes from the $\beta$-delayed $\gamma$ decay of $^{20}$Mg, and the 1634~keV $\gamma$-ray comes from the $\beta$-delayed $\gamma$ decay of the daughter nucleus $^{20}$Na. The 1042~keV $\gamma$-ray comes from the $\beta$-delayed $\gamma$ decay of $^{18}$Ne, which is a main contaminant in the secondary beam. The 511~keV $\gamma$-ray comes from the positron-electron annihilation. Besides, there are two $\gamma$-rays from natural background, i.e., the 1461~keV $\gamma$-ray from the $^{40}$K decay \cite{Cameron_NDS2004} and the 2614~keV $\gamma$-ray from the $^{208}$Tl decay \cite{Martin_NDS2007}. The $\beta$ decay branching ratio to the 984~keV state of $^{20}$Na is estimated to be 66.9(46)\% by using the counts of the 984~keV $\gamma$-ray. This value agrees fairly well with the literature value of 69.7(12)\% \cite{Piechaczek_NPA1995}. In Fig.~\ref{GammaSpec}(b), the 238~keV, 275~keV, 1233~keV and 1298~keV $\gamma$ lines correspond to the de-excitations from the four lowest excited states in $^{19}$Ne after proton emissions from the states in $^{20}$Na. In order to distinguish individual decay branches contained in each proton peak, it is necessary to conduct a proton-$\gamma$-ray coincidence analysis. An example of $\gamma$-ray spectrum in coincidence with $p4$ is shown in Fig.~\ref{p4GammaSpec}, the ratio of the efficiency corrected counts of 275~keV and 1233~keV $\gamma$ lines to the count of $p4$ can be used to estimate the branching ratio for this decay branch. If none of the four expected $\gamma$ lines were observed clearly in the proton-coincident $\gamma$-ray spectrum, this decay branch should be assigned as a proton emission to the ground state of $^{19}$Ne. A classification of the components contained in each proton peak is summarized in Table~\ref{L20Mg}, the decay branch is marked with a ``?" in the case which only one event is observed in the proton-coincident $\gamma$-ray spectrum and therefore more statistics are needed to give a clear identification of these questionable components. The latest proton-separation energy value of $^{20}$Na $S_p(^{20}\mathrm{Na})=2190.1(11)$~keV \cite{Wrede_PRC2010} is adopted in the determination of the excitation energies of the states in $^{20}$Na. With the above information obtained, together with the branching ratios for each proton peak presented in Table~\ref{E20Mg}, the corresponding branching ratios for the $^{20}$Na states populated in $^{20}$Mg decay can be estimated accordingly. Combined with the half-life and excitation energies measured in the present work, as well as the $Q_{\mathrm{EC}}=10626.9(22)$~keV extracted from the latest mass measurements of $^{20}$Na \cite{Wrede_PRC2010} and $^{20}$Mg \cite{Gallant_PRL2014}, the corresponding log~$ft$ values for each $^{20}$Na state can be calculated. As for the 2645~keV state in $^{20}$Na, no obvious proton peak around the expected energy of 455~keV can be observed in the particle spectrum presented in Fig.~\ref{ParticleSpec}. Only an upper limit of its branching ratio is estimated to be 0.24(3)\%, failing to further improve the limit value of 0.02\% \cite{Wallace_PLB2012}. The results are listed in Table~\ref{B20Mg} and Table~\ref{Log20Mg}. However, the reliability of the assignment of the proton emission suffers from the number of counts in the $p\gamma$ coincidence spectra to some extent. In general, our branching ratios values are slightly lower than the literature values due to the fact that not all the decay branches can be unambiguously identified in the proton-$\gamma$-ray coincidence analysis. The low detection efficiency for $\gamma$-ray is mainly responsible for the missing decay branch.

\begin{table}
\caption{\label{L20Mg}The decay branches contains in each proton peak and the corresponding initial states in $^{20}$Na and the final states in $^{19}$Ne.}
\begin{center}
\begin{ruledtabular}
\begin{tabular}{cc}
Proton peak   & $^{20}$Na level$\rightarrow^{19}$Ne level (keV) \\
    \hline
    $p1$    & $2998\rightarrow0$ \\
    $p2$    & $4801\rightarrow1536$ \\
    $p3$    & $5142\rightarrow1536?$ \\
    $p4$    & $3863\rightarrow0, 4130\rightarrow275$ \\
    $p5$    & $4130\rightarrow0, 4362\rightarrow275?, 5595\rightarrow1508$ \\
    $px$    & $4721\rightarrow275, 5982\rightarrow1536?$ \\
    $p6$    & $4801\rightarrow275$ \\
    $p7$    & $4801\rightarrow0$ \\
    $p8$    &  ?\\
    $p9$    & $6318\rightarrow275$ \\
    $p10$   & $6523\rightarrow238, 6523\rightarrow275$ \\
    $p11$   & $6523\rightarrow0$ \\
\end{tabular}
\end{ruledtabular}
\end{center}
\end{table}

\begin{table*}\footnotesize
\caption{\label{B20Mg}The excitation energies and decay branching ratios ($br$) for the states in $^{20}$Na.}
\begin{center}
\begin{ruledtabular}
\begin{tabular}{cccccccccccc}
    \multicolumn{2}{c}{S. Kubono \cite{Kubono_PRC1992}} & \multicolumn{2}{c}{J. G\"{o}rres \cite{Gorres_PRC1992}} & \multicolumn{2}{c}{A. Piechaczek \cite{Piechaczek_NPA1995}} & \multicolumn{2}{c}{J. P. Wallace \cite{Wallace_PLB2012}} & \multicolumn{2}{c}{M. V. Lund \cite{Lund_PhD2016}} & \multicolumn{2}{c}{Present Work} \\
    $E^*$ (keV) & $br$ (\%) & $E^*$ (keV) & $br$ (\%) & $E^*$ (keV) & $br$ (\%) & $E^*$ (keV) & $br$ (\%) & $E^*$ (keV) & $br$ (\%) & $E^*$ (keV) & $br$ (\%) \\
\hline
    990   & 85    &       & 74(7) & 984.25(10) & 69.7(12) &       &       &       &       & 983.9(22) & 66.9(46) \\
    2637  & $\leq1$   & 2645  & $\leq0.2$ & 2645  & $\leq0.1$ & 2647(3) & $\leq0.02$ &       &       & 2645  & $\leq0.24$ \\
    3046  & 9     & 3006(10) & 10.7(5) & 3001(2) & 11.5(14) &       &       & 2970(8) & 10.11(85) & 2998(13) & 8.6(7) \\
          &       &       &       &       &       & 3075(15) & 0.5(1) &       &       &       &  \\
    3868  & 5     & 3869(11) & 5.4(5) & 3874(15) & 4.8(6) &       &       & 3846(10) & 6.59(39) & 3863(14) & 3.7(4) \\
    4090  &       &       &       & 4123(16) & 2.7(3) &       &       & 4094(2) & 3.21(25) & 4130(22) & 2.3(5) \\
          &       &       &       &       &       &       &       &       &       & 4721(18) & 1.0(7) \\
          &       &       &       & $\sim4800$ & $\geq1.9$ &       &       & 4760(4) & 3.16(22) & 4801(32) & 1.2(4) \\
          &       &       &       &       &       &       &       & 5507(10) & 1.80(17) &       &  \\
          &       &       &       & $\sim5600$ & $\geq1.5$ &       &       & 5604(5) & 0.16(6) & 5595(17) & 0.7(3) \\
          &       &       &       &       &       &       &       & 5836(13) & 0.97(15) &       &  \\
          &       &       &       & 6266(30) & 1.2(1) &       &       & 6273(7) & 1.93(17) & 6318(17) & 1.6(9) \\
    6440  & 1.5   & 6533(15) & 3.0(8) & 6521(30) & 3.3(4) &       &       & 6496(3) & 4.16(20) & 6523(28) & 3.6(6) \\
          &       &       &       & $\sim6770$ & $\geq0.03$ &       &       & 6734(25) & 0.358(12) &       &  \\
          &       &       &       & $\sim6920$ & $\geq0.01$ &       &       &       &       &       &  \\
          &       &       &       &       &       &       &       & 7183(16) & 0.093(8) &       &  \\
          &       &       &       & $\sim7440$ & $\geq0.01$ &       &       &       &       &       &  \\
\end{tabular}
\end{ruledtabular}
\end{center}
\end{table*}

\begin{table*}\footnotesize
\caption{\label{Log20Mg}The log~$ft$ values for the states in $^{20}$Na.}
\begin{center}
\begin{ruledtabular}
\begin{tabular}{cccccccccccc}
    \multicolumn{2}{c}{S. Kubono \cite{Kubono_PRC1992}} & \multicolumn{2}{c}{J. G\"{o}rres \cite{Gorres_PRC1992}} & \multicolumn{2}{c}{A. Piechaczek \cite{Piechaczek_NPA1995}} & \multicolumn{2}{c}{J. P. Wallace \cite{Wallace_PLB2012}} & \multicolumn{2}{c}{M. V. Lund \cite{Lund_PhD2016}} & \multicolumn{2}{c}{Present Work} \\
    $E^*$ (keV) & log~$ft$ & $E^*$ (keV) & log~$ft$ & $E^*$ (keV) & log~$ft$ & $E^*$ (keV) & log~$ft$ & $ E^*$ (keV) & log~$ft$ & $E^*$ (keV) & log~$ft$ \\
\hline
          & 3.87  &       & 3.70(5) & 984.25(10) & 3.83(2) &       &       &       &       & 983.9(22) & 3.80(4) \\
    2637  & $\geq5.42$ & 2645  & $\geq5.85$ & 2645  & $\geq6.24$ & 2647(3) & $\geq6.9$ &       &       & 2645  & $\geq5.82$ \\
    3046  & 4.31  & 3006(10) & 3.99(4) & 3001(2) & 4.08(6) &       &       & 2970(8) & 4.10(8) & 2998(13) & 4.15(4) \\
          &       &       &       &       &       & 3075(15) & 5.41(9) &       &       &       &  \\
    3868  & 4.26  & 3869(11) & 3.99(5) & 3874(15) & 4.17(6) &       &       & 3846(10) & 4.11(6) & 3863(14) & 4.23(5) \\
    4090  &       &       &       & 4123(16) & 4.33(6) &       &       & 4094(2) & 4.33(8) & 4130(22) & 4.40(9) \\
          &       &       &       &       &       &       &       &       &       & 4721(18) & 4.5(3) \\
          &       &       &       & $\sim4800$ & $\leq4.23$ &       &       & 4760(4) & 4.08(7) & 4801(32) & 4.36(11) \\
          &       &       &       &       &       &       &       & 5507(10) & 3.99(9) &       &  \\
          &       &       &       & $\sim5600$ & $\leq3.97$ &       &       & 5604(5) & 5.00(38) & 5595(17) & 4.24(19) \\
          &       &       &       &       &       &       &       & 5836(13) & 4.09(15) &       &  \\
          &       &       &       & 6266(30) & 3.72(6) &       &       & 6273(7) & 3.55(9) & 6318(17) & 3.48(25) \\
    6440  & 3.68  & 6533(15) & 3.08(22) & 6521(30) & 3.13(6) &       &       & 6496(3) & 3.09(5) & 6523(28) & 3.01(8) \\
          &       &       &       & $\sim6770$ & $\leq5.01$ &       &       & 6734(25) & 4.00(3) &       &  \\
          &       &       &       & $\sim6920$ & $\leq5.39$ &       &       &       &       &       &  \\
          &       &       &       &       &       &       &       & 7183(16) & 5.14(67) &       &  \\
          &       &       &       & $\sim7440$ & $\leq4.99$ &       &       &       &       &       &  \\
\end{tabular}
\end{ruledtabular}
\end{center}
\end{table*}

A comparison between the mirror decays of $^{20}$Mg and $^{20}$O also provides opportunity to investigate the isospin asymmetry. The degree of isospin symmetry breaking can be reflected through the mirror asymmetry parameter $\delta=\frac{\mathrm{log}~ft^{+}}{\mathrm{log}~ft^{-}}-\mathrm{1}$, where the $\mathrm{log}~ft^{+}$ and $\mathrm{log}~ft^{-}$ values are associated with the $\beta^{+}$ decay of $^{20}$Mg and the $\beta^{-}$ decay of $^{20}$O, respectively \cite{Lund_PhD2016}. According to the compilation \cite{Tilley_NPA1998}, $Q_{\mathrm{\beta^{-}}}(^{20}\mathrm{O})=3814$~keV, hence only the two energetically accessible low-lying mirror transitions can be taken into consideration. In table~\ref{M20Mg}, the information of the mirror transitions extracted from the present measurement is summarized, and the large isospin asymmetry observed in the second mirror transitions confirms the conclusion reported in Ref.~\cite{Piechaczek_NPA1995}.

\begin{table*}
\caption{\label{M20Mg}Comparison between the transitions in the mirror $\beta$ decays of $^{20}$Mg and $^{20}$O.}
\begin{center}
\begin{ruledtabular}
\begin{tabular}{cccc}
    Transitions & log~$ft$ & Ref.  & $\delta$ \\
\hline
    $^{20}$O$\rightarrow^{20}$F 1057 keV & 3.740(6) & D. E. Alburger \cite{Alburger_PRC1987} &  \\
    $^{20}$Mg$\rightarrow^{20}$Na 984.25(10) keV & 3.83(2) & A. Piechaczek \cite{Piechaczek_NPA1995} & 0.024(6) \\
    $^{20}$Mg$\rightarrow^{20}$Na 983.9(22) keV & 3.80(4) & Present work & 0.016(11) \\
          &       &       &  \\
    $^{20}$O$\rightarrow^{20}$F 3488 keV & 3.65(6) & D. E. Alburger \cite{Alburger_PRC1987} & \\
    $^{20}$Mg$\rightarrow^{20}$Na 3001(2) keV & 4.08(6) & A. Piechaczek \cite{Piechaczek_NPA1995} & 0.12(3) \\
    $^{20}$Mg$\rightarrow^{20}$Na 2970(8) keV & 4.10(8) & M. V. Lund \cite{Lund_PhD2016} & 0.12(3) \\
    $^{20}$Mg$\rightarrow^{20}$Na 2998(13) keV & 4.15(4) & Present work & 0.14(3) \\
\end{tabular}
\end{ruledtabular}
\end{center}
\end{table*}

\section{Conclusion}
A detailed study of the $\beta$ decay of $^{20}$Mg was performed by using a detection system for charged-particle decay studies with a continuous-implantation method. A proton-$\gamma$-ray coincidence analysis was applied to the identification of $\beta$-delayed proton decay branches of $^{20}$Mg, and a new proton branch with an energy of 2256~keV was observed. The improved spectroscopic information on the decay property of $^{20}$Mg was deduced. The good agreement between our results with the literature values proves the validity of the described analysis method to obtain information about $\beta$ decay precisely. The isospin asymmetry for the mirror decays of $^{20}$Mg and $^{20}$O was investigated, as well. To clarify the remaining problems on the 2645~keV state in $^{20}$Na and construct the decay scheme of $^{20}$Mg completely, a further experiment with higher $\gamma$ detection efficiency and improved statistics is highly desirable on the basis of present work.


\begin{acknowledgments}
We acknowledge the continuous effort of the HIRFL staff for providing good-quality beams and ensuring compatibility of the electronics. We would like to thank Jun Su for the very helpful discussions. This work is supported by the National Basic Research Program of China under Grant No. 2013CB834404, and the National Natural Science Foundation of China under Grants No. 11375268, No. 11475263, No. U1432246, No. U1432127, No. 11505293 and No. 11635015.
\end{acknowledgments}



\end{document}